\documentclass[prd,twocolumn,floats,floatfix,nofootinbib]{revtex4}
\usepackage[dvips]{graphicx}
\usepackage{graphics}
\usepackage{dcolumn}
\usepackage{amssymb}
\usepackage{bm}
\usepackage{amsmath}

\def\spose#1{\hbox to 0pt{#1\hss}}

\def\lta{\mathrel{\spose{\lower 3pt\hbox{$\mathchar"218$}}
     \raise 2.0pt\hbox{$\mathchar"13C$}}}
\def\gta{\mathrel{\spose{\lower 3pt\hbox{$\mathchar"218$}}
     \raise 2.0pt\hbox{$\mathchar"13E$}}}
\newcommand{\be}{\begin{equation}}
\newcommand{\en}{\end{equation}}
\newcommand{\bea}{\begin{eqnarray}}
\newcommand{\ena}{\end{eqnarray}}

\begin{document}

\title{Gravitational wave recoils in non-axisymmetric Robinson-Trautman spacetimes}

\author{R. F. Aranha$^{1}$, I. Dami\~ao Soares$^{1}$ and E. V. Tonini$^{2}$}

\address{$^{1}$Centro Brasileiro de Pesquisas F\'isicas, Rio de Janeiro 22290-180, Brazil,\\
$^{3}$Instituto Federal do Esp\'irito Santo, Vit\'oria 29040-780, Brazil.}
\email{rfaranha@cbpf.br; ivano@cbpf.br;tonini@cefetes.br }

\date{\today}

\begin{abstract}
We examine the Bondi-Sachs momentum fluxes carried out by gravitational waves and the
associated net kick velocities in non-axisymmetric Robinson-Trautman spacetimes,
using characteristic initial data for the dynamics that correspond to the early
post-merger state of a non-head-on collision of two boosted black holes. We make
a parameter study of the kick distributions, corresponding to an extended range of the collision
angle parameter $\rho_0$ of the initial data. Typically for the equal mass case
the net gravitational wave momentum flux is nonzero, suggesting that these systems
might be a candidate to an approximate description of the post-merger phase of a
non-head-on collision of black holes not preceded by a premerger inspiral phase,
as for instance colliding black holes in unbounded trajectories.
For the range of $\rho_0$ examined ($3^{\circ} \leq \rho_0 \leq 110^{\circ}$)
we show that the kick distributions as a function of the symmetric mass parameter $\eta$
satisfy a $\eta$-scale law obtained from an empirical modification of the Fitchett law,
by properly including a parameter $C$ that accounts for the non-zero net gravitational
momentum wave fluxes for the equal mass case. This law fits accurately the kick distributions
for the range of $\rho_0$ examined, with a rms normalized error of the order of, or
smaller than $5 \%$. For $\rho_0=0^{\circ}$ (the head-on case) we have $C=1$ corresponding to
the Fitchett law. For the equal mass case we verified that the nonzero
net gravitational wave momentum flux increases as $\rho_0$ increases. The threshold
for this behavior is $\rho_0 \simeq 55^{\circ}$ beyond which it decreases continuously. For $\rho_0 \geq 50^{\circ}$
the distribution is a monotonous function of $\eta$. The maximum net kick velocity occurs for $\rho_0 \simeq 55^{\circ}$
at $\eta=0.25$, being of the order of $190 {\rm km/s}$ for the boost parameter considered.
The general angular patterns of the gravitational waves emitted are examined, for initial and later times.
Our analysis includes the two polarization modes of the wave zone curvature
that appear in the case of non-head-on collisions, connected to the two {\it news}
present in the radiative field.
\end{abstract}


\maketitle

\section{Introduction}

The collision and merger of two black holes is presently considered to be an important
astrophysical configuration where processes of generation and emission of gravitational waves take
place (cf. \cite{pretorius} and references therein). The radiative transfer involved in these processes,
evaluated in the full nonlinear regime of General Relativity, shows that gravitational waves extract mass,
momentum and angular momentum of the source, and may turn out to be fundamental for the astrophysics
of the collapse of stars and the formation of black holes. The process of momentum extraction and the
associated recoils in the system can have important consequences for astrophysical scenarios, as the
evolution and the population of massive black holes in galaxies or in the intergalactic
medium\cite{baker,merritt,favata}. Observational evidence of black hole recoils have been reported
in \cite{komossa} and references therein.
\par Gravitational wave recoils and the associated kick processes of two black holes have been
investigated within several approaches, most of them connected to binary black hole inspirals.
Post-Newtonian approximations (cf. \cite{blanchet1} and references therein) estimated the
kick velocity accumulated during the adiabatical inspiral of the system
plus the kick velocity accumulated during the plunge phase.
Sopuerta et al.\cite{sopuerta} computed the recoil velocity based on the close limit
approximation (CLA) supplemented with post-Newtonian (PN) calculations.
The first full numerical relativity (NR) evaluation of the recoil in nonspinning black hole
binaries was reported by Baker et al.\cite{baker} for a mass ratio $\simeq 0.667$,
while Gonz\'alez et al.\cite{gonzalez0} and Campanelli et al.\cite{camp1} simultaneously
obtained much larger recoils for black hole binaries with antialigned spins.
Gon\-z\'alez et al.\cite{gonzalez} undertook a more complete NR treatment
of kicks in the merger of black hole binaries by contemplating a larger
parameter domain.
For the case of small mass ratios in the interval $0.01 \leq \alpha \leq 0.1$ full
numerical relativity evaluations bridged with perturbative techniques were implemented in Refs.
\cite{gonzalez1,camp3,camp4,camp2}.
Le Tiec et al.\cite{blanchet}, combining PN+CLA methods, recently evaluated the
gravitational wave recoil in black hole binaries and showed that the ringdown phase
produces a significant anti-kick.
In the same vein Choi et al.\cite{choi} examined recoils in head-on collisions of
black holes, considering the head-on case as a model problem which can be seen as an
approximation to the final plunge to merger
and allow to isolate kick effects from the orbital inspiral motion.
Finally Rezzola et al.\cite
{rezzolla,rezzolla1} obtained an important injective relation
between the kick velocities and the effective curvature parameter of the global apparent horizon
in head-on collisions, using initial data derived in \cite{aranha1,aranhaT}.
In spite of the enormous progress achieved until now using approximation
methods and numerical techniques, the information on wave form patterns and radiative transfer processes in the
dynamics of gravitational wave emission is far from being complete\cite{pfeifer}.
\par In this paper we examine the distribution of kicks for a large domain of the collision
angle $\rho_0$ , a parameter of the initial data used. The work completes Ref. \cite{aranha11}
where we examined the energy and momentum extraction in the post-merger phase of a non-head-on
collision of two black holes, in the realm of Robinson-Trautman (RT) spacetimes\cite{rt}.
Our treatment is based on the Bondi-Sachs (BS) four- momentum conservation laws\cite{bondi, sachs, sachs1,cqg}
that regulate the radiative transfer processes involved in the emission of gravitational waves.
The characteristic initial data constructed for the dynamics already present a global apparent horizon
so that the dynamics covers the post-merger phase of the system up to the final configuration of the remnant black hole.
Due to the presence of a global apparent horizon the initial data  effectively
represents an initial single distorted black hole which is evolved via the RT dynamics.
Similarly to the case of the CLA -- where the perturbation equations of a black
hole\cite{zerilli,teuk} are feeded either with numerically generated, or with Misner, or Bowen-York-type initial
data -- we feed the (nonlinear) RT equation with the above mentioned characteristic data.
It is in this sense that we denote the dynamics thus generated as ``the post-merger phase
of two colliding black holes''. The interpretation of the outcomes of the RT dynamics
should be considered with the above caveats.
In the remaining of the introduction we give a brief review of the
basic properties of RT spacetimes which will be necessary for the discussions in the paper.

\par RT spacetimes\cite{rt} are asymptotically flat solutions of Einstein's vacuum equations
that describe the exterior gravitational field of a bounded system
radiating gravitational waves. In a suitable coordinate system the metric can be expressed as
{
\begin{eqnarray}
\label{eq1}
\nonumber
ds^2&=&\Big({\lambda(u,\theta,\phi)- \frac{2 m_{0}}{r}+ 2 r
\frac{{K}_{,u}}{K}}\Big) du^2 +2du dr\\
&-&r^{2}K^{2}(u,\theta,\phi)~ \Big( d \theta^{2}+\sin^{2}\theta d \phi^{2} \Big),
\end{eqnarray}}
where
{
\begin{eqnarray}
\nonumber
\lambda(u,\theta,\phi)=\frac{1}{K^2}&-&\frac{(K_{,\theta}~ \sin
\theta/K)_{,\theta}}{K^2
\sin \theta} \\
&+&\frac{1}{\sin^2 \theta}\left(\frac{K_{,\phi}^2}{K^4}-\frac{K_{,\phi
\phi}}{K^3} \right).
\label{eq2}
\end{eqnarray}}
The Einstein vacuum equations for (\ref{eq1}) result in

{
\begin{eqnarray}
\label{eq3}
-6 m_{0}\frac{{K}_{,u}}{K}+\frac{1}{2 K^2}\Big(\frac{(\lambda_{,\theta} \sin
\theta)_{,\theta}}{ \sin \theta}+\frac{\lambda_{,\phi \phi}}{ \sin^2 \theta}\Big)=0.
\end{eqnarray}}

\noindent Subscripts $u$, $\theta$ and $\phi$, preceded by a comma, denote
derivatives with respect to $u$, $\theta$ and $\phi$, respectively. $m_0 > 0$ is the only dimensional parameter of
the geometry, which fixes the mass and length scales of the spacetime. Eq. (\ref{eq3}), the RT equation,
governs the dynamics of the system and allows to evolve the initial data $K(u_0,\theta,\phi)$, given in the
characteristic surface $u=u_0$, for times $u > u_0$. For sufficiently regular initial data RT spacetimes exist globally
for all positive $u$ and converge asymptotically to the Schwarzschild metric as $u \rightarrow
\infty$\cite{chrusciel}.
Once the initial data $K(u_0,\theta,\phi)$ is specified, a unique apparent horizon (AH)
solution is fixed for that $u_0$\cite{AH}. Since the AH is the outer past marginally trapped
surface, the closest of a white hole definition (the remnant black hole will form as $u
\rightarrow \infty$), only the exterior and its future development via RT dynamics with outgoing gravitational
waves is of interest. We note that all the BS quantities, measured at the future null infinity
$\mathcal{J}^{+}$, are constructed and well defined under the outgoing radiation condition[17,18].
\par The field equations have a stationary solution that will play an important role in our
discussions,
{\small
\begin{equation}
K(\theta,\phi)=\frac{K_0}{\cosh \gamma+ ({\bf{n}} \cdot \hat{\bf{x}})\sinh \gamma},
\label{eq4}
\end{equation}}

\noindent where $\hat{\bf{x}}=(\sin \theta \cos \phi,\sin \theta \sin \phi, \cos \theta)$ is the unit
vector along an arbitrary direction ${\bf{x}}$, and ${\bf{n}}=(n_1,n_2,n_3)$ is a constant unit vector
(satisfying $n_{1}^{2}+n_{2}^{2}+n_{3}^{2}=1$). Also $K_0$ and $\gamma$ are constants. We note
that (\ref{eq4}) yields $\lambda=1/K_{0}^{2}$, showing its stationary character.
This solution can be interpreted\cite{bondi} as a boosted black hole along the axis determined by the unit vector
${\bf{n}}$ with boost parameter $\gamma$, or equivalently, with velocity parameter $v=\tanh \gamma$.
The Bondi mass function associated wit (\ref{eq4}) is $m(\theta,\phi)=m_{0}K^3(\theta,\phi)$ and the total
mass-energy of this gravitational configuration is given by the Bondi mass
{\small
\begin{eqnarray}
\label{eq5}
\nonumber
M_B&=&(1/4\pi)\int^{2\pi}_{0} d \phi \int^{\pi}_{0}  m(\theta,\phi)\sin \theta ~d\theta\\
&=&m_{0} K_{0}^{3}\cosh \gamma = m_{0} K_{0}^{3}/ \sqrt{1-v^2}.
\end{eqnarray}}
The interpretation of (\ref{eq4}) as a boosted black hole is relative to
the asymptotic Lorentz frame which is the rest frame of the black hole when $\gamma=0$.
\par In the paper we use units such that $8 \pi G=c=1$; $c$ is however restored in the definition of the kick velocity.
Except where explicitly stated, all the numerical results were done for $\gamma=0.5$. In our computational work we used $m_0=10$
but the results are given in terms of $u/m_0$. We should note that we can always set $m_0=1$ in the RT equation (\ref{eq3})
by the transformation $u \rightarrow {\tilde u}=u/m_0$.

\section{The Bondi-Sachs Four Momentum for RT Spacetimes and the Initial Data}

Since RT spacetimes describe asymptotically flat radiating spacetimes and the initial data of
its dynamics are prescribed on null characteristic surfaces, they are in the the realm of the 2+2
Bondi-Sachs formulation of gravitational waves in General Relativity\cite{bondi,sachs,sachs1}.
Consequently we must use suitable physical quantities of this formulation appearing in the description of
gravitational wave emission processes, as the BS four-momentum and its conservation laws. A
detailed derivation of the BS four-momentum conservation laws in RT spacetimes was given
in \cite{cqg}. We can show that, from the supplementary vacuum Einstein equations in the B-S integration scheme
together with the outgoing radiation condition, the B-S four momentum conservation laws for RT spacetimes are
{\small
\begin{eqnarray}
\label{eq12}
\frac{d P^{\mu}(u)}{d u}= P_{W}^{\mu}(u).
\end{eqnarray}}
In the above
{\small
\begin{eqnarray}
\label{eq9}
P^{\mu}(u)= \frac{1}{4 \pi} \int^{2 \pi}_{0} d \phi \int^{\pi}_{0} m(u,\theta,\phi)~l^{\mu} \sin \theta~ d \theta~~
\end{eqnarray}}
is the Bondi-Sachs four-momentum, where $m(u,\theta,\phi)$ is the Bondi mass function. The four vector
$l^{\mu}=(1, -\sin \theta \cos \varphi,-\sin \theta \sin \varphi, -\cos \theta)$, defined relative to an
asymptotic Lorentz frame, characterizes the
generators $l^{\mu} \Big(\partial/\partial U \Big)$ of the BMS translations in the
temporal and Cartesian $x,y,z$ directions of the asymptotic Lorentz frame\cite{sachs1}, and
{\small
\begin{eqnarray}
\label{eq11}
P_{W}^{\mu}(u)= -\frac{1}{4 \pi} \int^{2 \pi}_{0} d \phi \int^{\pi}_{0} K~l^{\mu} \Big( {c_u^{(1)}}^2 + {c_u^{(2)}}^2 \Big) \sin \theta~ d \theta~~
\end{eqnarray}}
is the net flux of energy-momentum carried out by the the gravitational waves. In (\ref{eq11}), the quantities ${c_u^{(1)}}$ and ${c_u^{(2)}}$
are the {\it news} functions for RT spacetimes expressed as
{\small
\begin{eqnarray}
\label{eq7}
\nonumber
c_{,u}^{(1)}(u,\theta,\phi)&=&\frac{1}{2} \Big( {\mathcal P}_{,\theta \theta}-{\mathcal
P}_{,\theta} \cot \theta -\frac{{\mathcal P}_{,\phi\phi}}{\sin^2 \theta}\Big),\\
c_{,u}^{(2)}(u,\theta,\phi)&=&\frac{1}{\sin \theta} \Big({\mathcal P}_{,\theta \phi}-
{\mathcal P}_{,\phi} \cot \theta\Big),
\end{eqnarray}}
where we have introduced the variable ${\mathcal P} \equiv 1/K$, for notation convenience.
We remark that $c^{(1)}_{,u}=0=c^{(2)}_{,u}$ for the boosted Schwarzschild solution
(\ref{eq4}), as should be expected.
The mass-energy conservation law (Eq. (\ref{eq12}) for $\mu=0$) is the Bondi mass formula.
Our main interest here is the analysis of the momentum conservation, (Eq. (\ref{eq12}) for $\mu=x,y,z$.
Due to the planar nature of a general collision, namely, the motion of the
two initial colliding black holes and the motion of the remnant are restricted to a plane, without
loss of generality we will fix this plane as the $(x,z)$-plane so that the momentum conservation equations
relevant to our discussion reduce to
{\small
\begin{eqnarray}
\label{eq13}
\frac{d{\bf P}(u)}{d u}={\bf P}_{W}(u),
\end{eqnarray}}
where ${\bf P}_W(u)=(P^{x}_{W}(u),0,P^{z}_{W}(u))$, with
{\small
\begin{eqnarray}
\label{eq15-i}
{P}_{W}^{x}(u)&=&\frac{1}{4 \pi}  \int^{2 \pi}_{0}  d \phi \int^{\pi}_{0} \sin^2 \theta \cos \phi~ K \Big( {c_u^{(1)}}^2 + {c_u^{(2)}}^2 \Big) d \theta,~~~~~~\\
\label{eq15-ii}
{P}_{W}^{z}(u)&=&\frac{1}{4 \pi}  \int^{2 \pi}_{0} d \phi \int^{\pi}_{0} \cos \theta \sin \theta~ K \Big( {c_u^{(1)}}^2 + {c_u^{(2)}}^2 \Big) d \theta.~~~~\\
\nonumber
\end{eqnarray}}
Obviously $P^{y}$ is conserved, a consequence of ${P}_{W}^{y}(u)=0$ for all $u$.
\par The initial data to be used was derived in Ref. \cite{aranha11} and is interpreted as representing two instantaneously
colliding Schwarzchild black holes in the $(x,z)$ plane, at $u=u_0$,
{\small
\begin{eqnarray}
\nonumber
&&K(u_0,\theta,\phi)=\Big(\frac{\alpha_1}{\sqrt{{\cosh \gamma+ \cos \theta \sinh \gamma}}}+\\
&&\frac{\alpha_2}{\sqrt{{\cosh \gamma - (\cos \rho_0 ~\cos \theta+ \sin \rho_0~ \sin\theta \cos \phi)\sinh \gamma}}}\Big)^2.~~~~~
\label{eq16}
\end{eqnarray}
}
In the derivation of (\ref{eq16}) it turns out that $\alpha_2/\alpha_1$ is the mass ratio of the Schwarzschild mass
of the initial data, as seen by an asymptotic observer. It is also worth mentioning the following properties of (\ref{eq16}):
(i) for $\alpha_2=0$ $(\alpha_1 \neq 0)$ or $\alpha_1=0$ $(\alpha_2 \neq 0)$ the initial data (\ref{eq16}) corresponds
to a boosted Schwarzschild black hole along, respectively, the positive $z$-axis or along the direction of the
unit vector ${\bf n}=(-\cos \rho_0,0,\sin \rho_0)$ with respect to an asymptotic Lorentz frame (cf. eq. (\ref{eq4})).
(ii) The specific combination (\ref{eq16}) of two black hole solutions (\ref{eq4}) is not arbitrary but arises
as the conformal factor of an asymptotically flat 3-geometry which is a solution of the constraint $^{(3)}R=0$.
(iii) Additionally (\ref{eq16}) results in a planar dynamics, namely, for all $u$ the net gravitational wave
momentum flux ${\bf P}_W(u)$ is restricted to the plane determined by the unit vectors ${\bf n}_z$ and ${\bf n}$
defining the direction of motion of the two black holes entering in (\ref{eq16}); the momentum of the remnant black hole
is also contained in this plane, as should be expected. The above properties reinforces the interpretation of (\ref{eq16})
as related to the post-merger phase of a black hole collision. The parameter $\rho_0$ of the
initial data, which we denote the incidence angle, defines the direction of the second initial black hole with respect to the $z$-axis
of the asymptotic Lorentz observer. $\gamma$ is the boost parameter of the black hole solutions (\ref{eq4}) which enter
in (\ref{eq16}). As in the 1+3 numerical relativity approach the interpretation of the initial data parameters involves an
approximation, namely, that the initial gravitational interaction in the data is neglected.
\par As mentioned already this data has a single apparent horizon so that the evolution covers the post-merger
regime up to the final configuration, when the gravitational wave emission ceases. It is worth remarking here
that, in the full Bondi-Sachs problem, further data (the {\it news} functions) are needed to determine the
evolution of the system. However for the RT dynamics the {\it news} are specified once $K(u,\theta,\phi)$ is
given, cf. (\ref{eq7}).
\section{Numerical evolution}
The initial data (\ref{eq16}) is evolved numerically via the RT equation (\ref{eq3}), which is integrated using a
Galerkin method with a spherical harmonics projection basis space \cite{fletcher} adapted to the
non-axisymmetric dynamics of RT spacetimes. The implementation of the Galerkin method,
as well as its accuracy and stability for long time runs, is described in detail in Section
V of Ref. \cite{aranha11}. The autonomous dynamical system derived with the Galerkin basis projection
is integrated using a fourth-order Runge-Kutta recursive method (adapted to our constraints)
together with a C++ integrator\cite{aranha11} for a truncation $N=7$.
Exhaustive numerical experiments show that after a sufficiently long time $u \sim u_f$
all the modal coefficients of the Galerkin expansion become constant up to twelve significant digits, corresponding to
the final time of computation $u_f$. At $u_f$ the gravitational wave emission is considered to effectively cease.
By reconstructing numerically $K(u,\theta,\phi)$ for all $u>u_0$ we can obtain the time behavior of important
physical quantities, as for instance the net gravitational wave flux and the associated total impulse imparted
to the merged system by the emission of gravitational waves. From the final constant modal coefficients we obtain
$K(u_f,\theta,\phi)$ that, in all cases, can be approximated as
{\small
\begin{eqnarray}
\label{eq17}
K(u_f,\theta,\phi) \simeq \frac{K_{f}}{\cosh \gamma_{f}+ (n_{1f} ~\sin \theta \cos \phi+n_{3f}~ \cos \theta) \sinh \gamma_{f}}.~~
\end{eqnarray}
}
With the final parameters $(K_f,\gamma_f,n_{1f},n_{3f})$ obtained from the final
modal coefficients, we have in all cases that the rms error of Eq. (\ref{eq17}) is of the order of, or smaller than $10^{-12}$.
The final configuration corresponds then to a Schwarzschild black hole
(cf. (\ref{eq4})) along the direction determined by ${\bf{n}}_f=(n_{1f},0,n_{3f})$, with a final boost parameter $\gamma_{f}$
and a final Bondi rest mass $m_0 K_{f}^{3}$. In all cases $\gamma_{f}< \gamma$ and $K_f > 1$.
The angle $\rho_f=\cos^{-1}(n_{3f})$ defines the direction of the remnant with respect to the $z$-axis.
Within the numerical error of our computation we have $(n_{1f})^2+(n_{3f})^2=1$ as expected.
\par The values of the parameters of the remnant black hole are one of the basic results
to be extracted from our numerical experiments, and are included in the Tables of the next Sections.

\section{Gravitational wave net momentum fluxes and kicks in a non-head-on collision}

We can now examine the processes of momentum extraction and the associated impulses imparted
to the merged system by the emission of gravitational waves. Our starting point is the construction,
of the curves of the net momentum fluxes carried out by gravitational waves, via the numerically
integrated function $K(u,\theta,\phi)$. Our numerical work in the present paper contemplates the
parameter intervals $\alpha=(0,1]$ and $\rho_0=[3^{\circ},125^{\circ}]$, with $\gamma=0.5$ fixed.
\par Integrating in time the conservation Eq. (\ref{eq13}) we find that
{\small
\begin{eqnarray}
\label{eq18}
{\bf P}(u)- {\bf P}(u_0)={\bf I}_{W}(u),
\end{eqnarray}}
where
{\small
\begin{eqnarray}
\label{eq20-i}
{\bf I}_{W}(u)=\frac{1}{4 \pi} \int^{u}_{u_0} d u^{\prime}  \int^{2 \pi}_{0} d \phi
\int^{\pi}_{0} K \Big( {c_{u^{\prime}}^{(1)}}^2 + {c_{u^{\prime}}^{(2)}}^2 \Big){\bf {\hat n}}  \sin \theta d \theta~~
\end{eqnarray}}
is the impulse imparted to the merged system due to the momentum carried out by the
gravitational waves emitted up to the time $u$, with ${\bf {\hat n}}=(\sin \theta \cos \phi,0,\cos \theta)$.
\par
\begin{figure*}
\begin{center}
{\includegraphics*[height=6.45cm,width=8.1cm]{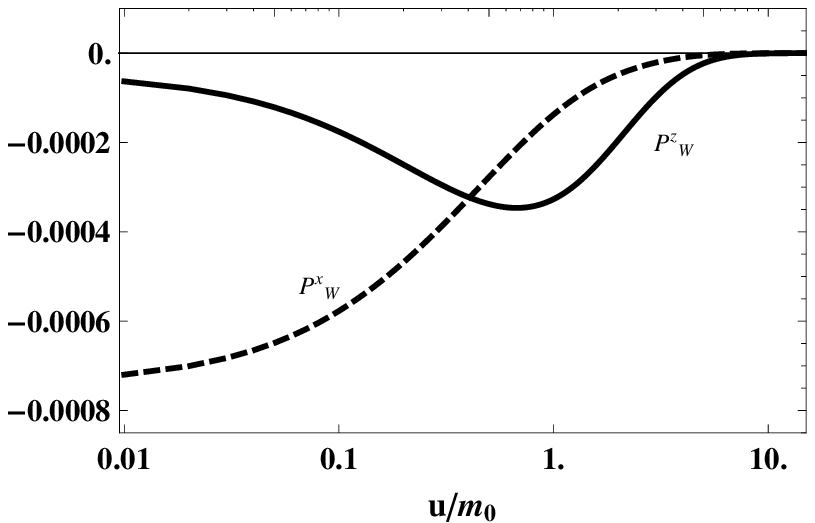}}{\includegraphics*[height=6.0cm,width=8.1cm]{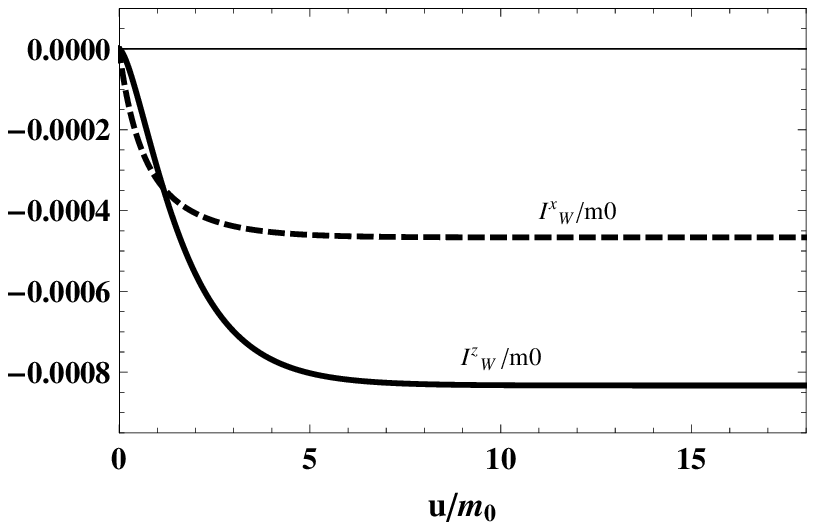}}
\caption{(left) Linear-log plot of the net fluxes of momentum $P_{W}^{x}(u)$ and $P_{W}^{z}(u)$ for $\alpha=0.25$,
$\gamma=0.5$ and incidence angle $\rho_0=60^{\circ}$. The figures shows a dominant deceleration regime
corresponding to a short pulse of gravitational bremsstrahlung of duration $\Delta u/m_0 \sim 10$.
(right) Plot of the gravitational wave impulses
$I_{W}^{x}(u)$ (dashed curve) and $I_{W}^{z}(u)$ (continuous curve) associated with the fluxes of the left figure.}
\label{master}
\end{center}
\end{figure*}

 In Fig. \ref{master}(left) we show the curves of the net momentum fluxes $P_{W}^{x}(u)$ and $P_{W}^{z}(u)$
for the mass ratio $\alpha=0.25$ and the incidence angle $\rho_0=60^{\circ}$, for $u>u_0$. For this value of $\rho_0$
the net momentum flux is negative for all $u$, corresponding to a strong deceleration regime of the system by
the emission of gravitational waves up to the final configuration of the remnant black hole, when the
gravitational wave emission ceases. We can also see that these fluxes correspond to short pulses of
gravitational bremsstrahlung in an interval $\Delta u/m_0 \sim 10 $. For incidence angles smaller than $55^{\circ}$
(and $\gamma=0.5$) the net momentum flux $P_{W}^{z}(u)$ is always positive for a short initial period.

\begin{table*}
\caption{Summary of our numerical results corresponding to an incidence angle of $\rho_0=21^{o}$
and boost parameter $\gamma=0.5$. }
\label{Table1}
\begin{tabular*}{\textwidth}{@{\extracolsep{\fill}}crrcrrrcrr@{}}
\hline
$\alpha$ & \multicolumn{1}{c}{$\eta$} & \multicolumn{1}{c}{$K_f$} & \multicolumn{1}{c}{$v_f/c=\tanh \gamma_f$} & \multicolumn{1}{c}{$\rho_f$}
& \multicolumn{1}{c}{$-I_{W}^{x}(u_f)/m_0$} & \multicolumn{1}{c}{$-V_{k}^{x}$} & \multicolumn{1}{c}{$I_{W}^{z}(u_f)/m_0$} & \multicolumn{1}{c}{$-V_{k}^{z}$} & $V_{k}~({\rm km/s})$ \\
\hline
$0.025$ & $0.0238$ &$1.045113$ &$0.446426$&  $0.37^{o}$ & $1.2792 \times 10^{-6}$&0.3362 &$8.1563\times 10^{-6}$&$2.1435$ & $2.1700$ \\
$0.050$ & $0.0453$ &$1.091454$ &$0.430867$& $0.77^{o}$ & $5.6819 \times 10^{-6}$&$1.3110$ &$3.3361\times 10^{-5}$ &$7.6973$& $7.8082$ \\
$0.100 $ & $0.0826$ & $1.187836$&$0.400255$& $1.65^{o}$ & $2.7635 \times 10^{-5}$&$4.9467$ &$1.3834 \times 10^{-4}$&$24.7624$&$25.2517$ \\
$0.150$& $0.1134$ & $1.289162$ &$0.370480$& $2.64^{o}$ & $7.4349 \times 10^{-5} $&$10.4105$&$3.1932 \times 10^{-4}$&$44.7116$ & $45.9076$ \\
$0.200$&$0.1388$& $1.395446$& $0.341701$  &$3.77^{o}$ &$1.5575 \times 10^{-4}$ & $17.1953$& $5.7681 \times 10^{-4}$& $63.6818$& $65.9625$  \\
$0.250$ &$0.1600$& $1.506701$& $0.314042$& $5.06^{o}$& $2.8314 \times 10^{-4}$& $24.8334$ & $9.0763\times 10^{-4}$& $79.6062$& $83.3897$  \\
$0.300$& $0.1775$&$1.622933$& $0.287600$& $6.53^{o}$& $4.6910 \times 10^{-4}$ & $32.9220$& $1.3051 \times 10^{-3}$ & $91.5906$& $97.3278$ \\
$0.400$& $0.2040$& $1.870356$& $0.238634$& $10.17^{o}$& $1.0733 \times 10^{-3}$&$49.2111$ & $2.2569 \times 10^{-3}$&$103.4813$& $114.5867$ \\
$0.500$& $0.2222$& $2.137755$& $0.195219$& $15.06^{o}$& $2.0929 \times 10^{-3}$&$64.2673$&$3.3174 \times 10^{-3}$&$101.8707$ & $120.4489$  \\
$0.525$& $0.2257$& $2.207669$& $0.185282$& $16.53^{o}$& $2.4287 \times 10^{-3}$&$67.7154$&$3.5817 \times 10^{-3}$&$99.8653$ & $120.6584$  \\
$0.600$& $0.2344$& $2.425151$& $0.157731$& $21.77^{o}$& $3.6700 \times 10^{-3}$&$77.1919$ & $4.3295 \times 10^{-3}$&$91.0621$& $119.3771$  \\
$0.700$ &$0.2422$& $2.732560$& $0.126756$& $31.14^{o}$& $5.9634 \times 10^{-3}$&$87.6814$ &$5.0986 \times 10^{-3}$ & $74.9656$& $115.3597$ \\
$0.800$& $0.2469$ &$3.059989$& $0.103331$& $44.09^{o}$ &$9.1477 \times 10^{-3}$&$95.7801$ &$5.3968 \times 10^{-3}$ &$56.5069$ & $111.2063$ \\
$0.900$&$0.2493$ &$3.407441$& $0.0888800$& $60.85^{o}$& $1.3413 \times 10^{-2}$&$101.7102$& $4.9651 \times 10^{-3}$& $37.6501$ & $108.4551$ \\
$1.000$ &$0.2500$& $3.774920$& $0.084214$& $79.50^{o}$& $1.8965 \times 10^{-2}$&$105.7664$& $3.5148 \times 10^{-3}$& $19.6022$ & $107.5675$ \\
\hline
\end{tabular*}
\end{table*}
\begin{table*}
\caption{Summary of our numerical results corresponding an incidence angle $\rho_0=60^{\circ}$
and boost parameter $\gamma=0.5$.}
\label{Table2}
\begin{tabular*}{\textwidth}{@{\extracolsep{\fill}}crrcrrrcrr@{}}
\hline
$\alpha$ & \multicolumn{1}{c}{$\eta$} & \multicolumn{1}{c}{$K_f$} & \multicolumn{1}{c}{$v_f/c=\tanh \gamma_f$} & \multicolumn{1}{c}{$\rho_f$}
& \multicolumn{1}{c}{$-I_{W}^{z}(u_f)/m_0$} & \multicolumn{1}{c}{$-V_{k}^{z}$} & \multicolumn{1}{c}{$I_{W}^{x}(u_f)/m_0$} & \multicolumn{1}{c}{$-V_{k}^{x}$} & $V_{k}~({\rm km/s})$ \\
\hline
$0.025$ & $0.0238$ &$1.046192$ & $0.449147$ & $0.96^{\circ}$ & $6.6062 \times 10^{-5}$& 1.7308 & $2.1448 \times 10^{-5}$ & $ 0.5619$ & $1.82$ \\
$0.050$ & $0.0453$ &$1.093620$ &$0.436498$ & $1.96^{\circ}$ & $2.7298 \times 10^{-4}$&$ 6.2611$ &$9.5087 \times 10^{-5}$ &$2.1809$& $6.63$ \\
$0.100 $ & $0.0826$ & $1.192193$&$0.412238$ & $4.09^{\circ}$ & $1.1585 \times 10^{-3}$&$ 20.5097$ &$4.6058 \times 10^{-4}$&$8.1544$&$22.07$ \\
$0.150$& $0.1134$ & $1.295730$ &$0.389466$& $6.39^{\circ}$ & $ 2.7465 \times 10^{-3} $&$37.8757$& $1.2341 \times 10^{-3}$&$17.0181$ & $41.52$ \\
$0.200$&$0.1388$& $1.404241$&$0.368273$ & $8.86^{\circ}$ & $5.1147 \times 10^{-3}$ & $55.4133$& $2.5750 \times 10^{-3}$& $27.8986$& $62.04$  \\
$0.250$ &$0.1600$& $1.517733$&$0.348716$ & $11.50^{\circ}$& $8.3286 \times 10^{-3}$& $71.4675$ & $4.6642\times 10^{-3}$& $40.0231$& $81.91$  \\
$0.300$& $0.1775$&$1.636210$&$0.330825$ & $14.30^{\circ}$& $1.2443 \times 10^{-2}$ & $85.2179$& $7.7020 \times 10^{-3}$ & $52.7483$& $100.22$ \\
$0.400$& $0.2041$& $1.888139$&$0.300047$& $20.36^{\circ}$& $2.3543 \times 10^{-2}$&$104.9265$ & $1.7522 \times 10^{-2}$&$78.0913$& $130.80$ \\
$0.500$& $0.2222$& $2.160553$&$0.275760$ & $26.92^{\circ}$& $3.8669 \times 10^{-2}$ & $115.1038$ & $3.4014 \times 10^{-2}$ & $101.2470$ & $153.30$  \\
$0.600$& $0.2344$& $2.451965$&$0.257534$& $33.79^{\circ}$ & $5.7953 \times 10^{-2}$&$117.9375$ & $5.9437 \times 10^{-2}$&$120.9574$& $168.94$  \\
$0.900$&$0.2493$ &$3.447766$&$0.232319$& $54.03^{\circ}$ & $14.0634 \times 10^{-2}$&$102.9436$& $21.5876 \times 10^{-2}$& $158.0197$ & $188.59$ \\
$1.000$&$0.25$ &$3.819731 $&$0.231060 $& $60.00^{\circ}$ &$17.5950 \times 10^{-2}$  & $94.7135$  & $30.47557 \times 10^{-2}$ &$164.0495$  & $189.43$  \\
\hline
\end{tabular*}
\end{table*}

\par The behavior of the associated impulses is illustrated in Fig. \ref{master} (right).
As expected $I_{W}^{z}(u)$ and $I_{W}^{x}(u)$ are negative for all $u$
and tend to a constant negative value (a plateau) for large $u \sim u_f$ corresponding to the
final configuration of the system, that of the remnant black hole. The plateau
is considered to be reached when $|{\bf I}_{W}(u)-{\bf I}_{W}(u+h)| \lesssim 10^{-10}$,
where $h$ is the stepsize of the integration used for the evaluation of ${\bf I}_{W}(u)$.
At this stage the remnant black hole has a momentum ${\bf P}=(n_{1f},0,n_{3f})~P_f$, with
{\small
\begin{eqnarray}
\label{momentumF}
P_f=m_0 K_{f}^{3} \sinh \gamma_f.
\end{eqnarray}}
The numerical Tables include values of the parameters characterizing the remnant black hole
for several $\rho_0$.
Typically the net total impulse imparted to the system has a dominant contribution from the deceleration regimes
(where $P_{W}^{z}(u)<0$ and $P_{W}^{x}(u)<0$) and will correspond to a net kick on the merged system. As we will discuss
later this net total impulse corresponds to the momentum of the remnant in a zero-initial-Bondi-momentum frame.
\par From Eq. (\ref{eq18}) we derive that
{\small
\begin{eqnarray}
\label{eq21}
{\bf P}(u_f)-{\bf P}(u_0)={\bf I}_{W}(u_f),
\end{eqnarray}
where the right-hand side of (\ref{eq21}) are the nonzero components of the net total impulse
${\bf I}_W(u_f)$ generated by the gravitational waves emitted. The values of ${\bf I}_W(u_f)$
correspond to the final plateaux which are present the impulse curves for any value of the
initial data parameters, as illustrated in Fig. \ref{master} (right).
\par We define the net kick velocity ${\bf V}_{k}$ as proportional to the net momentum imparted to the system by the total impulse
of the gravitational waves. This definition is based on the impulse function ${\bf I}_W(u)$
evaluated at $u=u_f$ (cf. eqs. (\ref{eq21})) and are in accordance with \cite{gonzalez}.
We obtain (restoring universal constants)
{\small
\begin{eqnarray}
\label{eq23-0}
{\bf V}_{k}&=&\frac{c}{m_0 K_{f}^{3}}~{\bf I}_{W}(u_f),
\end{eqnarray}}
with modulus
{\small
\begin{eqnarray}
\label{eq23}
V_{k}=\frac{c}{m_0 K_{f}^{3}}~\sqrt{(I_{W}^{x}(u_f))^2+(I_{W}^{z}(u_f))^2} ~,
\end{eqnarray}}
where $m_0 K_{f}^{3}$ is the rest mass of the remnant black hole. Taking into
account the momentum conservation equations evaluated at $u=u_f$ we interpret (\ref{eq23-0})
as the balance between the Bondi momentum of the system and the impulse of the gravitational waves
in a zero-initial-Bondi-momentum frame, which can then be compare with the results of the
literature. We remark that the zero-initial-Bondi- momentum frame is the inertial frame
related to the asymptotic Lorentz frame used in our computations by a velocity transformation with
velocity parameter ${\bf v}_B={\bf P}(u_0)/m_0 K_f^{3}$; in the parameter domain of our numerical experiments
the relativistic corrections in this transformation may be neglected.
We note that the velocity (\ref{eq23-0}) is directed along an axis
making the angle $\Theta_f=\arctan(I_{W}^{x}(u_f)/I_{W}^{z}(u_f))$ with the negative $z$-axis
of the zero-initial-Bondi-momentum frame.
\par For our initial data (\ref{eq16}) we have numerically evaluated
$V_k$ contemplating an extended range of the parameters $\alpha$ and $\rho_0$, with fixed $\gamma=0.5$.
The numerical results, illustrated in Tables \ref{Table1} and \ref{Table2} for the
cases $\rho_0=21^{\circ}$ and  $\rho_0=60^{\circ}$ respectively, are used to
construct the distribution curves of the kick velocities $V_{k}$ versus the symmetric mass parameter
$\eta=\alpha/(1+\alpha)^2$ In our numerical evaluations we contemplated fifteen values of $\rho_0$ in the
interval $3^{\circ}-125^{\circ}$.
These distributions are shown in Fig. \ref{fitchett}, for two separate domains of
the incidence angle $\rho_0$, the first corresponding to a domain of $\rho_0$ for which $V_k$ for $\alpha=1$
increases with $\rho_0$, and the latter for which $V_k$ for $\alpha=1$ decreases with the increase of $\rho_0$.
The threshold between the two behaviors is $\rho_0 \simeq 55^{\circ}$. The numerical evaluations
for the cases $\rho_0=110^{\circ},~115^{\circ},~125^{\circ}$ included in Table \ref{Table3} were not
not included to avoid the overcluttering in the Figures. The continuous curves are the
least-square-fit of the points to the empirical analytical formula
{\small
\begin{eqnarray}
\label{eq24}
V= A \eta^2 (1- 4 C \eta)^{1/2} (1+B \eta) \times 10^{3} ~{\rm km/s},
\end{eqnarray}}
with best-fit parameters given in Table \ref{Table3}.
\begin{table*}
\caption{Data for gravitational wave recoil of the equal-mass case ($\alpha=1$) and the best fit parameters for the
empirical law ~$V=A \eta ^2 (1- 4C \eta)^{1/2} (1+B \eta) \times 10^{3}~{\rm km/s}$. For the case $\alpha=1$ we
have the exact relation $\rho_f=(180^{\circ}-\rho_0)/2$.}
\label{Table3}
\begin{tabular*}{\textwidth}{@{\extracolsep{\fill}}crccrrr@{}}
\hline
$\rho_0$ & \multicolumn{1}{c}{$K_f$} & \multicolumn{1}{c}{$I_W(u_f)/m_0$ ($\alpha=1$)} & \multicolumn{1}{c}{$V_k(\alpha=1)~{\rm km/s}$} & \multicolumn{1}{c}{$A$} & \multicolumn{1}{c}{$B$} & {$C$}    \\
\hline
$3^{o}$ & $3.76843271$ &$0.00300016$ &$16.81843886$ &$3.63330391$&2.66808801 &$0.99802690$   \\
$10^{o}$ & $3.76980618$ &$0.00963759$ &$53.96754802$ &$3.63652172$&$2.62946623$ &$0.97947440$  \\
$21^{o} $ & $3.77491989$ & $0.01932698$&$107.78594166$ &$3.64355211$& $2.43737398$ &$0.91349987$  \\
$30^{o}$& $3.78172010$ & $0.02592443$ &$143.80115617$&$3.63128481$&$2.21967562$ &$0.83395393$  \\
$40^{o}$& $3.79190470$ & $0.03139079$ &$172.72343203 $& $3.56021379$&$1.95471187$ &$0.72811277$  \\
$45^{o}$&$3.79797695$& $0.03326940$&$182.15398438$ &$3.49052673$& $1.82788018$ &$0.67157657$   \\
$50^{o}$ &$3.80466278$& $0.03452600$&$188.06979866$&$3.40420169$ &$1.69568427$& $0.61463197$   \\
$55^{o}$& $3.81192716$&$0.03517381$&$190.50524979$ &$3.29095966$& $1.56847542$ &$0.55735444$  \\
$60^{o}$& $3.81973073$& $0.03521390$&$189.55581124$&$3.15252462$& $1.45291476$ &$0.50196512$  \\
$70^{o}$& $3.83677573$& $0.03357055$&$178.31193985$ &$2.81848260$& $1.20442961$ &$0.39473745$ \\
$80^{o}$& $3.855392222$& $0.02997635$&$156.92580163$ &$2.40820424$& $0.96981877$ &$0.29582448$  \\
$90^{o}$& $3.87510068$ & $0.02499139$& $128.84352941$&$1.94862882$&$0.76602544$& $0.21165881$   \\
$110^{o}$&$3.83677573$ & $0.01286913$& $68.35512596$ &$1.03429322$&$0.39544396$& $0.07400366$   \\
$115^{o}$& $3.82802972$ & $0.01027124 $& $54.93108403 $&$0.83429182 $&$0.31046537 $& $0.04432110 $   \\
$125^{o}$& $3.81192716$ & $0.00596900 $ & $32.32877651 $ &$0.49586913 $&$0.17357402  $& $0.00048779$   \\
\hline
\end{tabular*}
\end{table*}
Eq. (\ref{eq24}) is an empirical modification of the Fitchett law\cite{fitchett,blanchet1},
where the additional parameter $C$ was empirically introduced to account for the nonzero
net gravitational wave momentum flux in the non-head-on collision case with mass ratio $\alpha=1$,
and reduces to the Fitchett law for $C=1$. The Fitchett law was derived from post-Newtonian analysis
and used by a number of authors\cite{gonzalez,blanchet,fitchett} to adjust the distribution of
kick velocities in numerical relativity evaluations of the gravitational wave recoil in
merging binary inspirals of black holes and consistently yields a zero result for the equal
mass case. Therefore the results for kick distributions in the non-head-on case have
no connection with black hole binary inspirals, but rather possibly with two colliding black
holes in pre-merger unbounded trajectories, or in hyperbolic encounters of two nonspinning black
holes followed by a merger, the latter configuration recently discussed by Gold and Br\"ugmann\cite{gold}.
We mention that actually this modification introduced in (\ref{eq24}) is the only one
that works to produce an accurate fit of our results, with normalized rms error
of the order of, or smaller than $0.5\%$.
\begin{figure*}
\begin{center}
%\vspace{0.7cm}
{\includegraphics*[height=6.2cm,width=8.2cm]{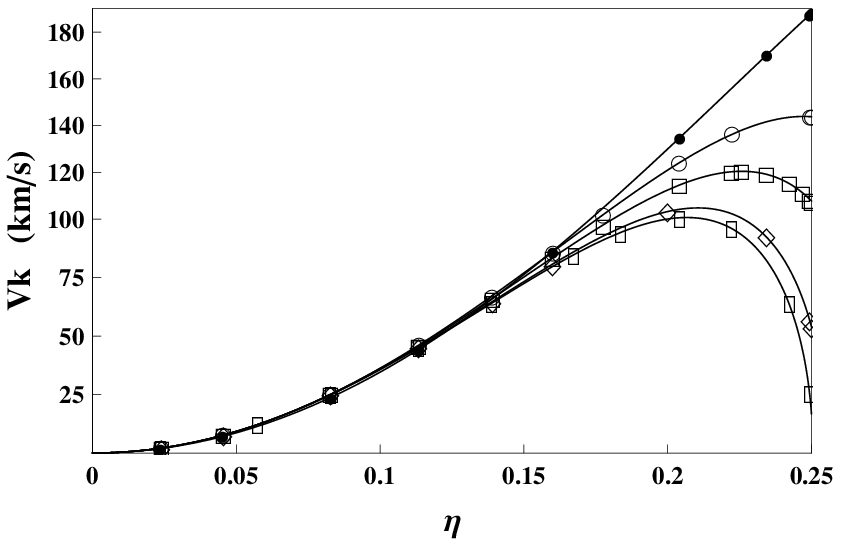}}~~~~{\includegraphics*[height=6.3cm,width=8.2cm]{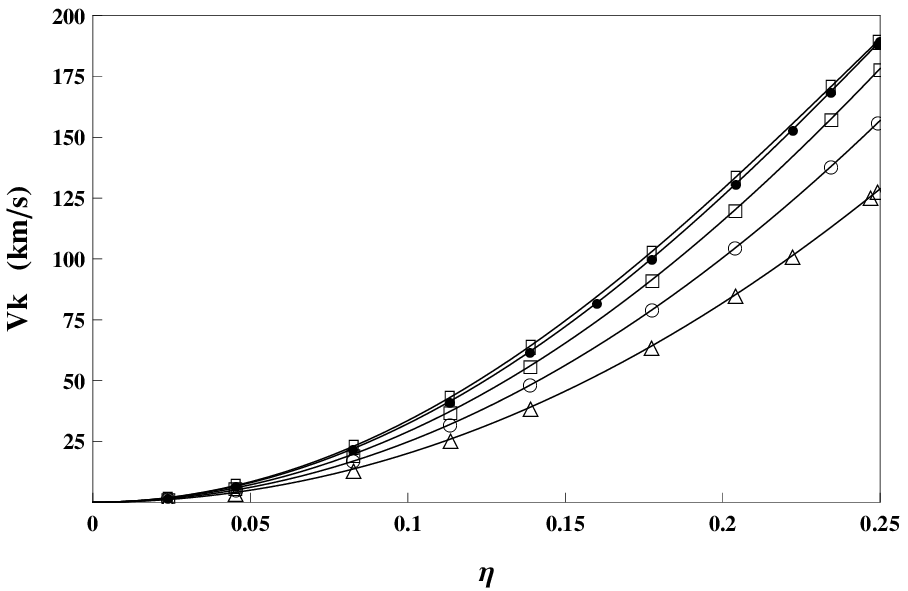}}
\caption{Plot of the points ($V_{k},\eta$), where $V_{k}$ is the net kick velocity (\ref{eq23})
due to the total impulse imparted on the merged system by the gravitational waves emitted: (left)
for $\rho_0=3^{\circ}$ (rectangles), $\rho_0=10^{\circ}$ (diamonds), $\rho_0=21^{\circ}$ (squares), $\rho_0=30^{\circ}$ (circles)
and $\rho_0=50^{\circ}$ (black dots); (right) for $\rho_0=55^{\circ}$ (rectangles), $\rho_0=60^{\circ}$ (black dots),
$\rho_0=70^{\circ}$ (squares), $\rho_0=80^{\circ}$ (circles) and $\rho_0=90^{\circ}$ (triangles).
The continuous curves of both figures correspond to the least-square-fit of the points to the analytical formula (\ref{eq24}),
with best fit parameters given in Table \ref{Table3}.
}
\label{fitchett}
\end{center}
\end{figure*}
\par The nonzero kick velocity for the non-head-on data with $\alpha=1$
deserves a further discussion that, without loss of generality, we
will restricted to the case $\rho_0=21^{\circ}$.
In this instance we can evaluate the components of the initial BS momentum
to be,
$P^{x}(0)/m_0 \simeq 4.489093$, $P^{z}(0)/m_0 \simeq 0.832004$ and
$P^{y}(0)/m_0 \simeq 0$,
with respect to an asymptotic Lorentz observer. This momentum vector, which lies in the
right quadrant of the upper hemisphere $z>0$ of the plane $x-z$, makes an angle
$\Theta_B=\arctan |P^{x}(0)/P^{z}(0)| \simeq 1.387537$ radians (or $\Theta_B \simeq 79.5^{\circ}$)
with the positive $z$-axis. This is also the direction of the nonzero momentum of the
remnant with respect to the same asymptotic Lorentz frame, determined by the angle $\rho_f$, which satisfies
$\rho_f=(180^{\circ}-\rho_0)/2$ for $\alpha=1$ and any $\rho_0$ (cf. Table \ref{Table3} and
Ref. \cite{aranha11}). The axis determined by $\rho_f \equiv \Theta_B$
actually plays an important role in the dynamics. If we take a new frame with its
$z$-axis coinciding with this axis the net gravitational wave momentum flux vector ${\bf P}_{W}(u)$ lies
along the new $z$-axis for all $u$. This was verified numerically by evaluating the ratios of the computed
fluxes, sampled in the interval $0 < u/m_0 \leq 890$, yielding in all cases
$\arctan|I_{W}^{x}(u)/I_{W}^{z}(u)| \simeq 1.387541$ with a relative error of the order
of $10^{-6}$. Still in this new frame the data will not be symmetric under $\theta \rightarrow \pi -
\theta$, leading to a nonzero net gravitational wave momentum flux, contrary to the case of
merging binary inspirals and head-on collisions. As expected the $z$-axis of the new frame
is the direction of the kick velocity since $\arctan |I_{W}^{x}(u_f)/I_{W}^{z}(u_f)|
\simeq 1.387547 ~{\rm rad}$ or $\simeq 79.5^{\circ}$ (within the precision of data in Table \ref{Table1}).
\par A remark is in order now concerning the balance between the total rest mass of the remnant and the
total net impulse of the gravitational waves in the distributions of the net kick velocity,
as observed from the numerical results displayed in the Tables. In the domain $0 < \rho_0 < 55^{\circ}$,
as $\eta$ increases from $0$ to $0.25$, both the parameter $K_{f}$ and the total net impulse {\small $I_{W}(u_f)=\sqrt{(I_{W}^{x}(u_f))^2+(I_{W}^{z}(u_f))^2}$} increase; however the increase of the
rescaled Bondi rest mass of the remnant, $K_{f}^{3}$, is smaller than the increase of
$I_{W}(u_f)$ up to $\eta \simeq 0.225$, implying that in this range the net kick velocity increases in accordance
with (\ref{eq23}). Beyond this point the increase of $K_{f}^{3}$ is larger than the increase of the total
net impulse leading to a decrease in the values of $V_k$ up to $\eta=0.25$. On the other hand, in the domain
$55^{\circ} < \rho_0 < 125^{\circ}$, the above behavior is reversed leading to the monotonous increase in
$\eta$ shown in Fig. \ref{fitchett}.
Finally we must comment that the parameter $C$ in the $\eta$-scaling law (\ref{eq24}), which assumes the value $C=1$
for head-on collisions and for merging black hole inspirals, decreases monotonically as $\rho_0$ increases, as can be seen from
Table \ref{Table3}.
\par In general, in a zero-initial-Bondi-momentum frame, the Bondi momentum of the merged
system satisfies ${\bf P}(u)={\bf I}_{W}(u)$ so that an integral curve ${\bf x}(u)$ of the wave
impulse vector field ${\bf I}_{W}(u)$, defined as $d{\bf x}/du={\bf P}(u)$, can give a schematic
picture of the motion of the merged system in this frame. In Fig. \ref{figCM} we
display this integral curve for $\alpha=0.2$ and $\rho_0=21^{\circ}$, generated with
initial conditions $x(u_0)=0=z(u_0)$ in the zero-initial-Bondi-momentum frame.
The initial phase of positive momentum flux along $z$ is responsible for the curved form
of the trajectory in the semiplane $z > 0$. For $u \rightarrow u_f$ the curve approaches
the asymptote with angle $\Theta_f=\arctan (I_{W}^{x}(u_f)/I_{W}^{z}(u_f)) \simeq 15.11^{\circ}$ with
respect to the negative $z$-axis of the zero-initial-Bondi-momentum frame, which is actually the
direction of the kick velocity in this frame.
In the case $\alpha=1$ the integral curve is a straight
line with angle $\Theta_f \simeq 79.5^{\circ} \equiv (180^{\circ}-21^{\circ})/2$ with
respect to the $z$-axis of the zero-initial-Bondi-momentum frame (cf. Table \ref{Table3}).
\begin{figure}[t]
\begin{center}
{\includegraphics*[height=5.5cm,width=8.5cm]{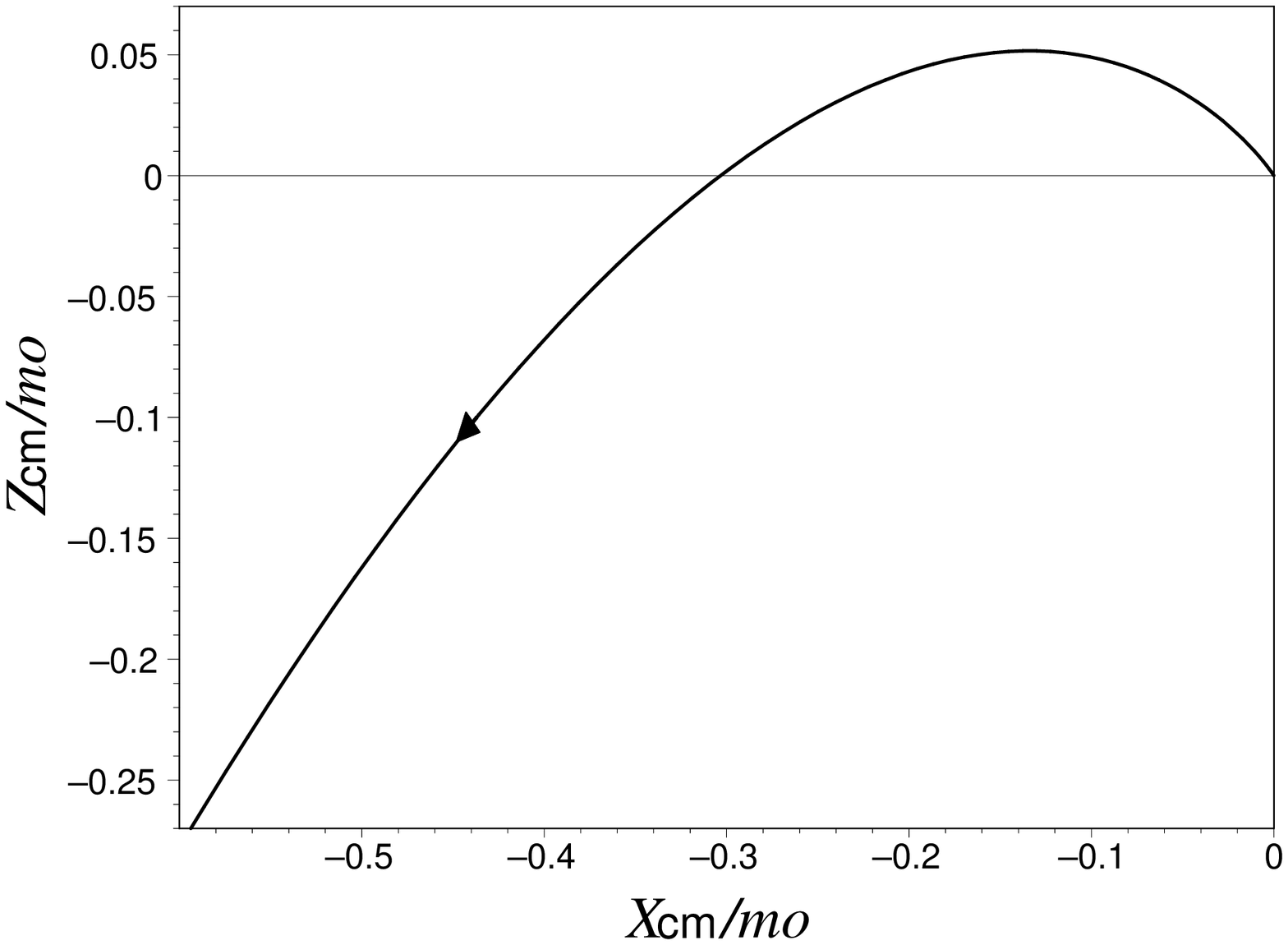}}
\caption{Plot of the integral curve ${\bf x}(u)$ of the wave impulse vector field ${\bf
I}_{W}(u)$
which can give a schematic picture of the motion of the system in the zero-momentum-Bondi
frame, for $\alpha=0.2$
and $\rho_0=21^{\circ}$. The asymptote of the curve as $u \rightarrow u_f$ makes an angle
$\Theta_f \simeq 15.11^{o}$
with the negative $z$-axis of this frame.}
\label{figCM}
\end{center}
\end{figure}

\section{The angular wave pattern and the bremsstrahlung regime of the gravitational waves}

The radiative character of RT spacetimes is given by the expression of its curvature
tensor that in a suitable semi-null tetrad basis\cite{aranha11} assumes the form
{\small
\begin{eqnarray}
\label{eq6N}
R_{ABCD}=\frac{N_{ABCD}}{r}+\frac{III_{ABCD}}{r^2}+\frac{II_{ABCD}}{r^3},
\end{eqnarray}}
where the quantities {\small $N_{ABCD}$, $III_{ABCD}$ and $II_{ABCD}$} are of the
algebraic type $N$, $III$ and $II$, respectively, in the Petrov classification
of the curvature tensor\cite{petrov}, and $r$ is the parameter distance along the
principal null direction $\partial /\partial r$. Eq. (\ref{eq6N}) displays the peeling
property\cite{peeling} of the curvature tensor, showing that indeed RT is
the exterior gravitational field of a bounded source emitting gravitational waves.
For large $r$ we have
{\small
\begin{equation}
R_{ABCD} \sim \frac{N_{ABCD}}{r},
\label{eq7N}
\end{equation}}
\noindent so that at large $r$ the gravitational field looks like a gravitational wave with propagation
vector $\partial /\partial r$. The nonvanishing of the {\small $N_{ABCD}$}
is therefore an invariant criterion for the presence of gravitational waves, and the
asymptotic region where ${\cal{O}}(1/r)$-terms are dominant defined as the wave zone.
The curvature tensor components in the above basis that contribute to {\small $N_{ABCD}$} are
{\small $R_{0303}=-R_{0202}=-D(u,\theta,\phi)/r+{\cal{O}}(1/r^2)$} and {\small $R_{0203}=-B(u,\theta,\phi)/r+{\cal{O}}(1/r^2)$}
where
{\small
\begin{eqnarray}
\nonumber
D(u,\theta,\phi)&=&-~P^2~ {\partial}_{u}~ \Big( \frac{c_{,u}^{(1)}(u,\theta,\phi)}{P}\Big),\\
B(u,\theta,\phi)&=&-~P^2 ~ {\partial}_{u}~ \Big( \frac{c_{,u}^{(2)}(u,\theta,\phi)}{P} \Big),
\label{eq27N}
\end{eqnarray}
}
with the {\it news} $c_{,u}^{(1)}(u,\theta,\phi)$ and $c_{,u}^{(2)}(u,\theta,\phi)$ given in (\ref{eq7}).
From (\ref{eq7N}) we can see that the functions $D$ and $B$ contain
all the information of the angular, and time dependence of the gravitational
wave amplitudes in the wave zone, once $K(u,\theta,\phi)$ is given. $D$ and $B$
actually correspond to the two polarization modes of the gravitational wave, transverse to its direction of
propagation in the wave zone. We note that in the axisymmetric case $B=0$, which is the case of
head-on collisions. We will consider the particular combination
{\small
\begin{eqnarray}
\label{penrose}
D(u,\theta,\phi)+i B(u,\theta,\phi) \sim ~(r \Psi_4),
\end{eqnarray}}
where $\Psi_4$ is the Weyl spinor associated with $N_{ABCD}/r$ in a suitable Newman-Penrose null tetrad basis\cite{penrose,aranhaTese}.
The quantity (\ref{penrose}) is specified once we
have the function $K(u,\theta,\phi)$, which is numerically obtained via the
numerical integration of the dynamics, as we have discussed.
\par In Fig. \ref{Plot1} we display the polar plots of $\sqrt{D^2+B^2}$ at early times $u=0.01$
({\rm dotted}),$u=0.05$ ({\rm dash-dotted}) and $u=0.1$ ({\rm continuous}), and initial data parameters
$\alpha=0.2$, $\rho_0=55^{o}$ and $\gamma=0.5$, with section by the plane $\phi=0^{o}$ (corresponding to the plane of collision $(x,z)$).
The plots in Fig. \ref{Plot1} (left) show, for each time, a pattern with two dominant lobes in the
forward direction of motion of the merged system. The direction of the Bondi momentum
vector at $u=0.01$ makes an angle $\Theta_B \simeq 8,43^{o}$ with the $z$ axis.
The pattern is typical of a bremsstrahlung process due to the deceleration of the merged system,
analogous to the electromagnetic bremsstrahlung of a charge decelerated along its direction of motion.
As time increases we observe that the cone enveloping the dominant lobes opens up and the amplitudes decrease. For later
times the pattern evolves to the expected quadrupole structure with a much smaller amplitude, as shown in Fig. \ref{Plot1} (right)
for $u=5.0$. We mention that the increase of the initial boost parameter $\gamma$ would sharpen the forward cone enveloping
of the two dominant lobes in the early regime, as expected in a ultrarelativistic configuration.
In our computations we fixed $m_0=10$. In the Figures the $z$ direction corresponds to the vertical axis.
\begin{figure*}[t]
\begin{center}
{\includegraphics*[height=7.8cm,width=7.5cm]{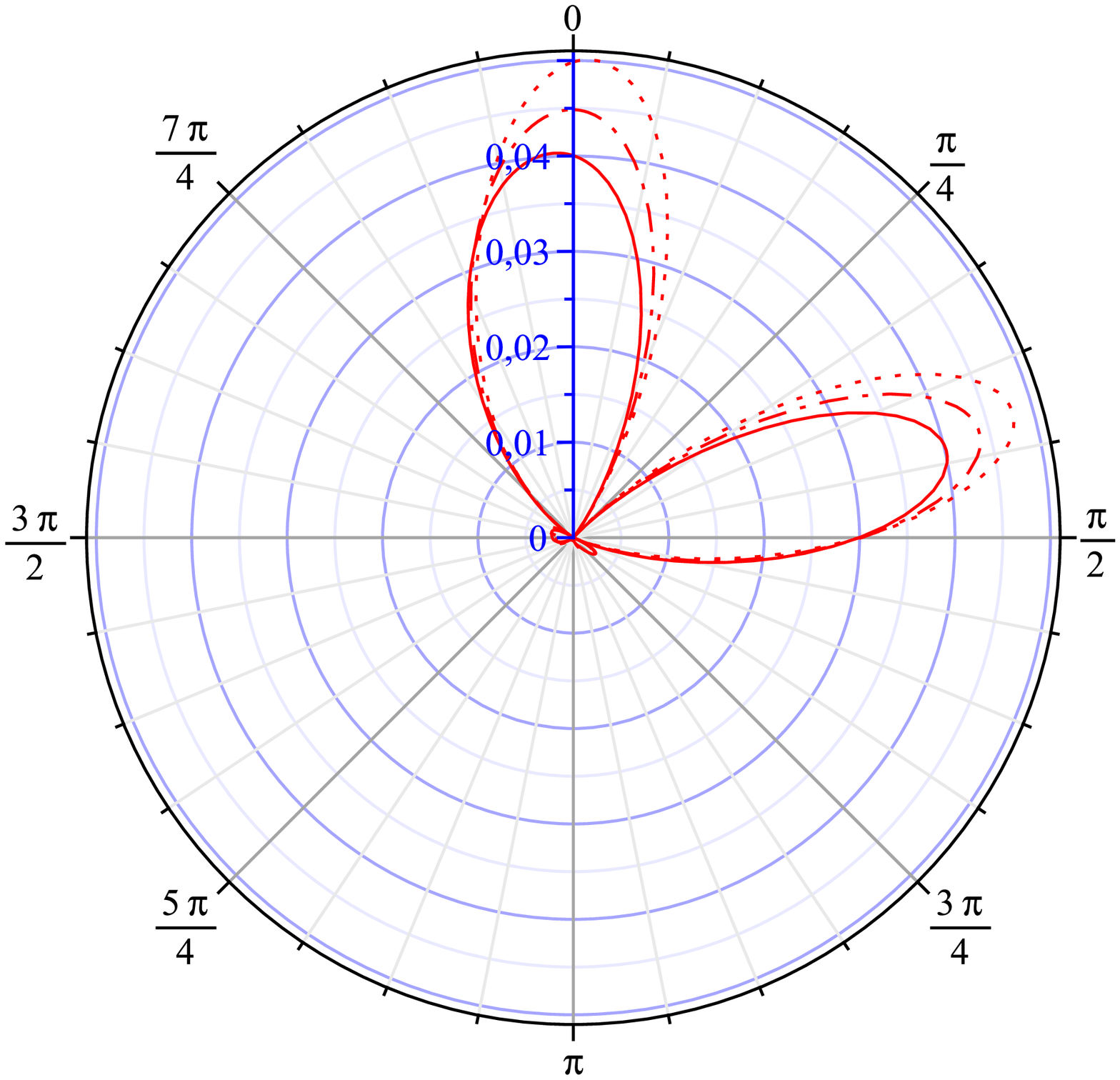}}{\includegraphics*[height=7.8cm,width=7.5cm]{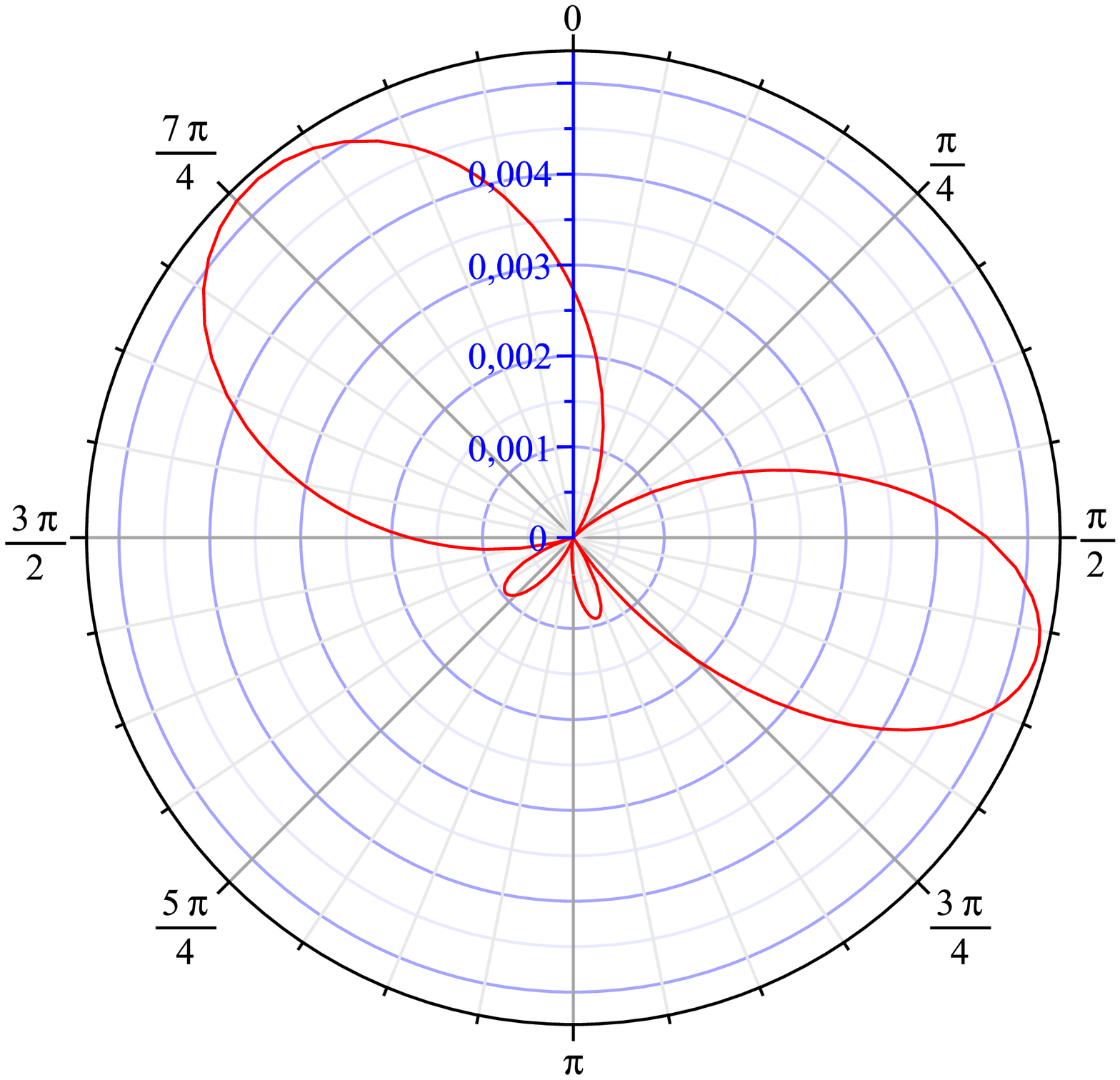}}
\caption{(Left) Polar plot of $\sqrt{D^2+B^2}$ (section by the plane $\phi=0^{o}$, corresponding to the plane of the collision) for times $u=0.01$ (dotted),
$u=0.05$ (dash-dotted) and $u=0.1$ (continuous), and initial data parameters
$\alpha=0.2$, $\rho_0=55^{o}$ and $\gamma=0.5$. The figure shows a typical bremsstrahlung pattern, corresponding to a
strong deceleration regime at early times, with two dominant lobes along the direction of motion of the merged system.
The cone enveloping the two dominant lobes opens up as $u$ increases. (Right) Polar plot of $\sqrt{D^2+B^2}$ for the same configuration
of the left figure, at a later time $u=5.0$, showing the opening of the lobes and the setting already of the final quadrupole pattern.
The $z$ direction corresponds to the vertical axis.}
\label{Plot1}
\end{center}
\end{figure*}
\begin{figure}
\begin{center}
{\includegraphics*[height=7.8cm,width=7.5cm]{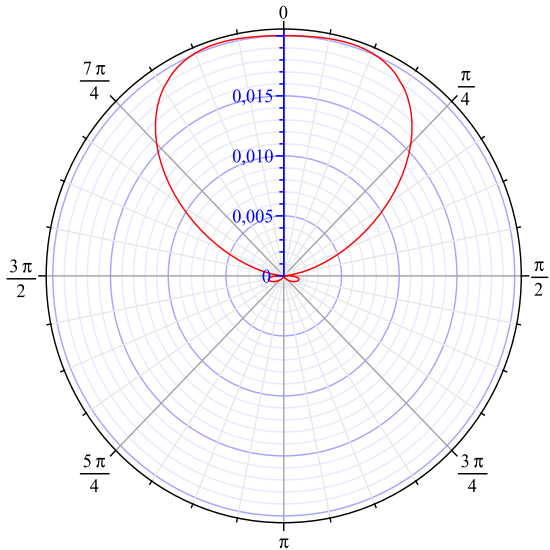}}
\caption{Polar plot of $\sqrt{D^2+B^2}$ (section by the plane $\phi=90^{o}$, corresponding to the plane $(y,z)$
orthogonal to the plane of collision) for a time $u=0.1$ and the same initial data parameters of previous figures.
The symmetry about the $z$-axis is in accordance with the conservation of $P_{W}^{y}(u)=0$.
Although gravitational waves are emitted outside the plane of collision, the zero net momentum flux
of this radiation component is consistent with the planar nature of the collision.}
\label{Plot2}
\end{center}
\end{figure}
\par In Fig. \ref{Plot2} we show the polar plot of $\sqrt{D^2+B^2}$, with section by the plane
$\phi=90^{o}$ (corresponding to the plane $y-z$, orthogonal to the plane of collision), at $u=0.1$. The same initial
data parameters of the previous Figures were used. As expected the pattern is symmetric about the $z$ axis
in accordance with the conservation of $P_{W}^{y}(u)=0$. We then see that, although gravitational waves are emitted
outside the plane of the collision, this radiation component has a zero net momentum flux. Therefore it does
not extract momentum of the system, consistent with the planar nature of the collision.

\section{Conclusions and Final Comments}

In the present paper we have examined the momentum extraction and the associated gravitational wave recoil
in the post-merger phase of two black holes in non-head-on collision, by contemplating an extended
domain of the incidence angle parameter $\rho_0$ present in the initial data. Our treatment is made in the
realm of Robinson-Trautman dynamics and based on the Bondi-Sachs characteristic formulation of
gravitational waves, and completes Ref. \cite{aranha11}. The net gravitational wave impulse associated with the
net momentum fluxes carried out by gravitational waves are evaluated, for an extended domain of the
incident angle parameter, $3^{\circ} \leq \rho_0 \leq 110^{\circ}$, of the initial data.
Typically the total net impulse is negative what, in terms of the Bondi-Sachs
momentum conservation laws, corresponds to a dominant deceleration regime
of the system due to the emission of gravitational waves. However, for relatively small values of
the mass ratio parameter $\alpha$ and of $\rho_0$, an initial positive impulse in the $z$ direction may be present,
which will be responsible for an initial inspiral branch in the motion of the system.
\par By using the Bondi-Sachs four momentum conservation laws we evaluate the net kick velocity $V_k$
imparted to the system as proportional to total net gravitational wave impulse in a
zero-initial-Bondi-momentum frame. For each value of the incidence angle $\rho_0$
considered we evaluate the distribution of $V_k$ as a function of the symmetric mass parameter $\eta$.
\par A novel feature of non head-on collisions ($\rho_0 \neq 0$) is the nonzero net gravitational wave fluxes
for the equal-mass case, contrary to the cases of head-on collisions and merging of black hole inspiral binaries.
This implies that the net kick velocity for non-head-on collisions are non-zero for the equal-mass case.
As a consequence the results for kick distributions in the non-head-on case have
no connection with black hole binary inspirals, but rather possibly with two colliding black
holes in pre-merger unbounded trajectories, or in hyperbolic encounters followed by a merger of two nonspinning black
holes as recently discussed by Gold and Br\"ugmann\cite{gold}.
\par These distributions of $V_k$ as a function of $\eta$ were shown to be fitted, for the whole domain of
$\rho_0$ considered, by the empirical law (\ref{eq24}) obtained by a modification of the Fitchett $\eta$-scaling law,
this modification corresponding to the introduction of an additional parameter $C$ to account for
the non-zero gravitational wave momentum fluxes in the equal mass case ($\eta=0.25$) of non-head-on collisions.
The best fit of the points ($V_k,\eta$) with the modified law is sufficiently accurate with a rms error
of the order of, or smaller than $0.5\%$ for all $\rho_0$. For $\rho_0=0^{\circ}$ (the case of head-on collisions)
we have $C=1$ and this distribution reduces to the Fitchett law, as expected.
For large incidence angles (e.g. $\rho_0 > 55^{\circ}$ in the case of $\gamma=0.5$) the distributions are
monotonous in $\eta$. We also verified that the best fit value of the parameter $C$ decreases, which is
$C=1$ for head-on collisions, decreases monotonically as the incidence angle $\rho_0$ increases.
\par We examined the behavior of the integral curves of the gravitational wave impulse ${\bf I}_W(u)$
that, in accordance with the Bondi-Sachs momentum conservation law, can describe the motion of the merged system
in a zero-initial-Bondi-momentum frame. This integral curve exhibits an initial inspiral branch in the positive
$z$ semi-plane whenever an initial phase of $P_{W}^{z}(u)>0$ is present.
\par Finally we have examined the angular patterns of the radiation both in the initial regime,
which is typically bremsstrahlung, and in larger times where the quadrupole pattern is already set up.

\par The authors acknowledge the partial financial support of CNPq/MCT-Brazil, through a Post-Doctoral Grant No. 201879/2010-7 (RFA),
Research Grant No. 306527/2009-0 (IDS), and of FAPES-ES-Brazil (EVT).

\end{document}